\documentstyle[aps,prd,preprint,tighten]{revtex}
\begin{document}
\preprint{\baselineskip 18pt{\vbox{\hbox{SU-4240-610} \hbox{June, 1995}}}}
%\vspace{35mm}
\title {Edge States in Canonical Gravity{\footnote{ Talk delivered by A.M. at
the XVIIth annual MRST Conference.}}}
\author{A.P. Balachandran, L. Chandar,  Arshad Momen}
\address{Department of Physics, Syracuse University,\\
Syracuse, NY 13244-1130, U.S.A.}
\date{}
\maketitle
\begin{abstract}
It is well-known that gauge fields defined on manifolds with spatial boundaries
support states localized at the boundary. In this talk, we show how similar
states arise in canonical gravity and discuss their physical relevance using
their analogy to quantum Hall effect.
\end{abstract}
%\vspace{5mm}

\newcommand{\be}{\begin{equation}}
\newcommand{\ee}{\end{equation}}
\newcommand{\bea}{\begin{eqnarray}}
\newcommand{\eea}{\end{eqnarray}}
\newcommand{\real}{{\rm l}\! {\rm R}}
\newcommand{\ra}{\rightarrow}
\newcommand{\tr}{{\rm tr}\;}
\baselineskip 15pt
\section{Introduction}

Recently there has been a renewal of interest in the problems of black hole
evaporation\cite{haw1} and the information loss puzzle.
Usually one studies quantum processes involving black holes
in a semiclassical approximation and typically one notices that
the situation calls for unknown physics involving
the event horizon and its surroundings \cite{susskind}.
There have been proposals to circumvent the situation by hypothesizing
 a stretched membrane with  certain classical properties, situated
just outside the black hole horizon,
which essentially  captures most of the important physics of black holes
\cite{susskind,membranebook,maggiore}. In a sense, this is a
phenomenological theory for black holes for an observer who is not falling into
the black hole.

The presence of such a membrane leads to an inner boundary for spatial slices.
We show that the presence of this boundary leads to an
{\it infinite} set of {\it
observables} which are completely
{\it localized at this boundary}.
These are obtained here in analogy
to  ``edge'' observables
in gauge theories defined on manifolds with boundaries \cite{sei,bal}.
Such observables have important physical relevance in many examples of
condensed matter physics, for instance,  the quantum Hall effect (QHE)
\cite{duality}. As described in a work under preparation \cite{us}, this
analogy with the quantum Hall effect can be useful for understanding
the origin of black hole entropy.

\section{ Existence of Edge States}

To show how edge states arise in gauge theories on manifolds,
we discuss a very simple example which we will also use later for QHE, namely
Chern-Simons theory for the $U(1)$ gauge group \cite{bal}. The action
describing Abelian
Chern-Simons theory is given by
\be
S_{cs} = \frac{k}{4 \pi} \int_{\cal M} A dA
\label{1.1}
\ee
where ${\cal M}$ is the space-time manifold having the topology of $ \real
\times D$, with $D$ being the disk. Note also that we are using differential
form
notation for the sake of brevity. As $A_0$ is nothing but a Lagrange multiplier
one immediately sees that the Gauss law constraint
 ( up to a numerical factor) is
\be
G(A) \equiv \epsilon^{ij} \partial_i A_j \approx 0.
\label{1.2}
\ee

Here it should be stressed that the existence of a Gauss law is the backbone
of the subsequent analysis.

In evaluating the Poisson brackets
of the constraints amongst themselves and for finding their action on the phase
space it is necessary to smear them with test functions
so that they become differentiable \cite{bal}. So, we smear the Gauss law with
the test function $\Lambda$,
\be
{\cal G}_{\Lambda} = \int_D \Lambda G(A) d^2x = \int_D \Lambda dA \approx 0.
\label{1.3}
\ee

Now we require that this functional generates gauge transformations, which in
turn requires that  $(\frac{\delta {\cal
G}_{\Lambda} }{ \delta A} )$ exists. However, one notes from (\ref{1.3}) that
\be
{\cal G}_{\Lambda} = \int_D A d\Lambda + \int_{\partial D} \Lambda A,
 \label{1.4}
\ee
and therefore differentiability requires the boundary condition
\be
\Lambda |_{\partial D} = 0.
\label{1.5}
\ee
Hence the Gauss law really is
\be
{\cal G}_{\Lambda} = \int_D A d\Lambda,  \label{1.6}
\ee
with the gauge parameter $\Lambda$ subjected to (\ref{1.5}).
Then,  due to the boundary condition (\ref{1.5}),
\be
\{ {\cal G}_{\Lambda}, {\cal G}_{\Lambda'}\} = \frac{2\pi}{k} \int_D \Lambda
d\Lambda' = 0 .
\label {1.7}
\ee

Recall that, any gauge invariant object is an observable and hence must have
zero Poisson bracket with the Gauss law. Thus we can define  the functional
\be
{\cal Q}(\xi) \equiv \int_D A d\,\xi
\label{1.8}
\ee
( which is inspired by the form (\ref{1.6}) ) with $\xi$ however
not subjected to the boundary condition (\ref{1.5}) , i.e.,
\be
\xi |_{\partial D}\;\; {\rm  not\; necessarily\; equal\; to\; 0}.
\label{1.9}
\ee

${\cal Q}(\xi)$ is an observable because
\be
\{ {\cal G}_{\Lambda} , {\cal Q}_(\xi) \} = \frac{2 \pi}{k} \int_{\partial D}
\Lambda d \xi = 0.
\label{1.10}
\ee
The fact that the observables ${\cal Q}(\xi)$ are really associated with
the edge can be shown as follows. If $\xi$ and $\xi'$ are
 the test functions such that they coincide
on the boundary so that
\be
\xi|_{\partial D} = \xi'|_{\partial D} \qquad \Rightarrow \qquad (\xi -
\xi')|_{\partial D} = 0,
\label{1.11}
\ee
then
\be
{\cal Q}(\xi) - {\cal Q}(\xi') = \int_D A d\,(\xi - \xi') = {\cal G}_{(\xi -
\xi')}.
\label{1.12}
\ee
So this weakly vanishes since ($\xi$ - $\xi'$) satisfies the condition
(\ref{1.5}) showing that ${\cal Q}(\xi)$ is localized at
the edge.

Finally  we see that these observables generate a $U(1)$ affine Lie algebra
at the edge,
\be
\{ {\cal Q}(\xi), {\cal Q}(\xi') \} = \frac{2 \pi}{k} \int_{\partial D} \xi \,
d\xi'
\label{1.13}
\ee

Now we can perform a similar analysis for the edge variables in
canonical gravity.

\section{ Edge States: A Case for Gravity}

To demonstrate the existence of edge states for gravity we will follow the
canonical treatment as before. The  standard ADM phase space analysis for
spacetimes foliated by Cauchy surfaces is discussed elsewhere in detail
\cite{wald,ash} and hence will not be repeated here. The constraints are
\bea
D_{a}p^{ab} &\approx &   0, \label{extra} \\
-{q^{\frac{1}{2}}}\;{^{(3)}R} + q^{-\frac{1}{2}}(p^{ab}p_{ab}
-\frac{p^{2}}{2}) &\approx &  0.
\label{9}
\eea
Here $p^{ab}$ is the momentum density conjugate to the spatial metric $q_{ab}$
and $D_a$ is the (projected)
covariant derivative compatible with $q_{ab}$.

As before, the vector constraint is to be smeared
with a form
$V_{a}$ that vanishes  at the boundaries  of the manifold while the scalar
constraint is to be  smeared with a test function $S$ that vanishes (along with
its derivatives) at the
boundaries. The boundaries here are the boundary $\partial {\cal B}_3$ of
${\cal B}_3$ (the spatial 3-ball whose boundary is the stretched membrane
enclosing the black hole ) and the
spatial infinity.

The smeared constraints are
\begin{eqnarray}
{\cal V}_{V}(q,p) & = & - 2 \int _{\Sigma} d^{3}x\; V_{a}D_{b}p^{ab} \approx 0,
\label{9.5}\\
{\cal S}_{S}(q,p) & = & \int _{\Sigma} d^{3}x\; S[-{q^{\frac{1}{2}}}\;{^{(3)}R}
+q^{-\frac{1}{2}}(p^{ab}p_{ab}-\frac{p^{2}}{2})] \approx 0, \label{10}
\eea
where
\bea
V_a |_{\partial \Sigma} = 0 \label{bc1} \\
S |_{\partial \Sigma} = 0, \qquad D_a S|_{\partial \Sigma}=0  \label{bc2}
\end{eqnarray}
The above conditions on the form $V_{a}$ and the function $S$ follow purely
from requiring differentiability in the phase space variables $q_{ab}$ and
$p^{ab}$ of (\ref{9.5}) and (\ref{10}).

The PB's  among the constraints are
\begin{eqnarray}
\{ {\cal V}_{V_{1}}\;,\;{\cal V}_{V_{2}}\} & = & {\cal V}_{[V_{1}\;,\;V_{2}]},
\nonumber \\
\{ {\cal V}_{V}\;,\;{\cal S}_{S}\} & = & {\cal S}_{{\cal L}_{V}S},\nonumber \\
\{ {\cal S}_{S_{1}}\;,\;{\cal S}_{S_{2}}\} & = & {\cal
V}_{S_{1}D\,S_{2}-S_{2}D\,S_{1}}. \label{11}
\end{eqnarray}

The construction of edge observables uses the trick that we have
already employed in the Chern-Simons theory.
We can construct  edge observables, analogous
to ${\cal S}_S$ and ${\cal V}_V$, whose  test
functions/forms will not be subjected to the boundary conditions (\ref{10}).
These observables turn out to
differentiable after {\em adding  suitable
surface terms}.
The difference of two of
these observables with different smearing forms/functions which coincide (along
with derivatives in the case of the latter) only
at the
boundaries is a constraint and hence they are truly edge degrees of freedom.

 To construct the edge observables arising from the vector ( diffeomorphism )
constraint  we
first  rewrite the vector constraint in (\ref{9.5}) after a partial
integration as
\be
{\cal V}_{V} =- \int _{\Sigma} d^{3}x\;q_{ab}{\cal L}_{V}p^{ab}.\label{redef}
\ee
In the above, let us replace $V$ by $W$ where $W$ is any vector field.
We require of $W$ that,
 at the boundaries of the manifold, it is
tangential to the boundary. Then it can be verified that the quantity so
obtained, namely
\be
{\cal D}_{W}=-\int_{\Sigma}d^{3}x\;q_{ab}{\cal L}_{W}p^{ab}.\label{diff}
\ee
continues to be differentiable in both $q_{ab}$ and $p^{ab}$.
It furthermore has weakly zero PB's with the constraints :
\begin{eqnarray}
\{ {\cal D}_{W}\;,\;{\cal V}_{V}\} & = & {\cal V}_{[W\;,\;V]},
\nonumber \\
\{ {\cal D}_{W}\;,\;{\cal S}_{S}\} & = & {\cal S}_{{\cal L}_{W}S}.\label{12}
\end{eqnarray}
The right hand sides in these equations are constraints and hence
weakly zero because
their respective test fields are easily verified
to satisfy the conditions (\ref{bc1}) and (\ref{bc2}).

The algebra of observables generated by ${\cal D}_W$ is seen to be
\be
\{ {\cal D}_{W_{1}}\;,\;{\cal D}_{W_{2}}\} = {\cal D}_{[W_{1}\;,\;W_{2}]}.
\label{13}
\ee

We are interested in observables which are supported
at the edge corresponding to the membrane rather than those which are
supported at spatial infinity. We will therefore hereafter assume that $W$ is
non-zero only at the
inner boundary and vanishes like $V$ at the boundary at infinity.

Next, let us look at the scalar constraint
${\cal S}_S$ :
\be
{\cal S}_{S}=\int _{\Sigma}d^{3}x\;
S[-{q^{\frac{1}{2}}}
\;{^{(3)}R}+q^{-\frac{1}{2}}(p^{ab}p_{ab}-\frac{p^{2}}{2})].\label{14}
\ee
The above is clearly differentiable in $p^{ab}$.
As for
differentiability in $q_{ab}$, it can be verified that a variation of
$q_{ab}$ induces surface terms in its variation. They vanish only if
the test functions $S$ satisfy (\ref{bc2}).
The condition on their derivatives emerges
because variation of ${^{(3)}R}$ contains second derivatives of the variation
of the metric $q_{ab}$ \cite{wald}. The boundary condition  in
(\ref{bc2}) on $S$ are in fact
got from this
requirement of differentiability of ${\cal S}_S$.

Consider (\ref{14}) with $S$ replaced by $T$,
which however does not have to satisfy the boundary
conditions satisfied by $S$. The only term in the expression that
would require careful scrutiny for differentiability in $q_{ab}$ is
\be
\int _{\Sigma}d^{3}x\; T[-{q^{\frac{1}{2}}}\; {^{(3)}R}].
\label{curvat}
\ee
The change in above term due to a variation $\delta q_{ab}$ is \cite{wald}
\be
-\int _{\Sigma}d^{3}x\; Tq^{\frac{1}{2}}[\frac{1}{2}\; {^{(3)}R}\;{q^{ab}}
- {^{(3)}
R^{ab}}]\delta q_{ab}-\int
_{\Sigma}d^{3}x\; T q^{\frac{1}{2}}[D^{a}D^{b}(\delta q_{ab})-D
^{a}(q^{cd}D_{a}\delta q_{cd})]. \label{varia}
\ee
Since the second term above contains derivatives of $\delta q_{ab}$,
(\ref{curvat}) is not differentiable with respect to $q_{ab}$.

Suppose  now that
\begin{eqnarray}
&&\delta q_{ab}|_{\partial \Sigma}=0, \label{(a)}\\
&&D_{a}T|_{\partial \Sigma}=0.
\label{(b)}
\end{eqnarray}
[ Note that (\ref{(b)}) implies that $T$ at the boundary
goes to a constant which can be non-zero.]
The terms involving derivatives of $\delta q_{ab}$ in
(\ref{varia}) give rise to surface terms in the variation. These surface
terms are now exactly cancelled by the
variation of
 \[ -2\int _{\partial \Sigma}T{\cal K}\sqrt{h} \]
where ${\cal K}_{ab}$ and $h_{ab}$ are respectively the extrinsic curvature and
the induced metric of the boundary $\partial \Sigma$ \cite{wald}.
Thus so long as the
conditions (\ref{(a)}) and (\ref{(b)}) above are met, we can define an edge
observable of the form
\be
{\cal H}_{T}=\int _{\Sigma}d^{3}x\; T[-{q^{\frac{1}{2}}}\;{^{(3)}
R} +q^{-\frac{1}{2}}(p^{ab}p_{ab}-
\frac{p^{2}}{2})]-2\int _{\partial \Sigma}d^{2}x\; h^{\frac{1}{2}}T{\cal K}.
\label{edH}
\ee

These edge observables, as presented, are independent of the observables
defined in the bulk. It is then not clear how coarse-graining over the edge
degrees of freedom can lead to an entanglement entropy for black holes
\cite{entropy}. We thus require a coupling between the edge and bulk degrees
of freedom. Quantum Hall effect again provides us with the model where such a
coupling occurs.  This is what we discuss in the next section.

\section{The Quantum Hall Effect: A Model for the
Dynamics of Edge Degrees of Freedom}

A simple effective action that describes the physics of quantum Hall effect
is the Chern-Simons action
added on to the usual electromagnetic action:
\bea
S_{bulk}&=& \int_{\cal M} d^3x \left[ -\frac{t}{4}F_{\mu \nu}F^{\mu \nu} -
\frac{\sigma_H}{2} \epsilon^{\mu \nu
\lambda}\,\,A_\mu \partial_\nu
A_\lambda \right], \label{MCS} \\
F_{\mu \nu}&=& \partial_\mu A_\nu - \partial_\nu A_\mu. \nonumber
\eea
Here $t$ is a constant related to the ``effective
thickness" of the Hall sample, while our metric is $(-1,+1,+1)_{diagonal}$.
The $\sigma _{H}$ that appears as the coefficient of the Chern-Simons term
is the Hall conductivity.

The connection of the above system with edge observables is also well-known.
The latter arise when we confine the above theory to a finite geometry (as is
appropriate for any
physical Hall sample).  From very general arguments first articulated
by Halperin
\cite{halperin}, the existence of chiral edge currents at the boundary can then
be established.

Naively, the theory in the bulk described by the action (\ref{MCS})
does not communicate with the theory describing
these chiral currents at the edge. It is then not clear how these edge
currents can have any role in the description of bulk phenomena.
However, gauge invariance \cite{duality,anomaly} allows us to put them
together.
 Thus the action (\ref{MCS}) under the gauge
transformation $A \ra A + d\alpha$
changes by the surface term
\be
-\frac{\sigma_H}{2}\int_{\partial {\cal M}} d\alpha \wedge A.
\label{st}
\ee
But, physics is gauge invariant. Therefore  it must be that there is a theory
at the
boundary describing the chiral edge currents which  is also gauge
non-invariant such
that the total action ($S_{tot}=S_{bulk}+S_{edge}$) is itself gauge invariant.
This line of argument \cite{duality} then leads us to the  action
\be
S_{tot} = S_{bulk} +\frac{\sigma_H}{2}
\int_{\partial {\cal M}} d\phi \wedge A -
\frac{\sigma_H}{4}
\int_{\partial {\cal M}} {\rm D}_\mu\phi\;{\rm D}^\mu \phi,  \label{totS}
\ee
\be
{\rm D}_\mu \phi = \partial_\mu\phi - A_\mu.  \nonumber
\ee

The field $\phi$ under a gauge transformation transforms as
\be
\phi \ra  \phi + \alpha  \label{gtphi} \qquad \Rightarrow
\qquad {\rm D}\phi \equiv d\phi - A \ra {\rm D}\phi \label{codep}
\ee
so that
\be
S_{tot} \ra S_{tot}.
\label{gttot}
\ee
The second term in (\ref{totS}) is the term which restores gauge invariance.
The last term is a kinetic energy term and is required if the
theory at the edge is to give rise to a {\em chiral} theory.

The dynamics of the edge field on the boundary and its coupling to the gauge
field allows one to calculate the entanglement entropy\cite{entropy}
arising due to a coarse-graining over the edge degrees of freedom
\cite{usnew}. One finds that the entropy scales as the perimeter of the disk.
So it is natural to inquire whether the black hole entropy also arises due to
a coarse-graining of the black hole edge states. However, to do this one
needs to know the dynamics of the black hole edge states.

\section{Edge Dynamics for Gravity?}

The ideas described in the previous sections give us hints about the dynamics
of the edge degrees of freedom for black holes. In fact, for (2+1) dimensional
gravity , which happens to be a Chern-Simons theory, one can find the edge
action exactly \cite{carlip,us}. However, the situation for the (3+1)
dimensional case is not so clear. One can write a kinetic energy term
for the edge degrees of freedom though their coupling to the external fields
would remain arbitrary.
We hope to report on these matters in detail some time in the future.

\noindent
\section*{Acknowledgements}
This work was supported by the US Department of Energy
contract number DE-FG02-85ER40231 and
by a Syracuse University Graduate fellowship awarded to A.M.

\end{document}